# Synergistic Drug Combination Prediction by Integrating Multi-omics Data in Deep Learning Models

Tianyu Zhang[1,4], Liwei Zhang[4], Philip R.O. Payne[1,3], Fuhai Li[1,2]

[1]Institute for Informatics (I2), [2]Department of Pediatrics, [3]Department of Medicine, Washington University School of Medicine, Washington University in St. Louis, St. Louis, Missouri, United States; [4]Dalian University of Technology, Dalian, China.
Corresponding author: Fuhai.Li@wustl.edu

**Abstract**

Drug resistance is still a major challenge in cancer therapy. Drug combination is expected to overcome drug resistance. However, the number of possible drug combinations is enormous, and thus it is infeasible to experimentally screen all effective drug combinations considering the limited resources. Therefore, computational models to predict and prioritize effective drug combinations is important for combinatory therapy discovery in cancer. In this study, we proposed a novel deep learning model, AuDNNsynergy, to prediction drug combinations by integrating multi-omics data and chemical structure data. In specific, three autoencoders were trained using the gene expression, copy number and genetic mutation data of all tumor samples from The Cancer Genome Atlas. Then the physicochemical properties of drugs combined with the output of the three autoencoders, characterizing the individual cancer cell-lines, were used as the input of a deep neural network that predicts the synergy value of given pair-wise drug combinations against the specific cancer cell-lines. The comparison results showed the proposed AuDNNsynergy model outperforms four state-of-art approaches, namely DeepSynergy, Gradient Boosting Machines, Random Forests, and Elastic Nets. Moreover, we conducted the interpretation analysis of the deep learning model to investigate potential vital genetic predictors and the underlying mechanism of synergistic drug combinations on specific cancer cell-lines.

# 1.Introduction

Combinatory therapies are being adopted in treating complicated diseases, e.g., cancer [1] and AIDS [2], which are more effective with reduced toxicity, and can overcome intrinsic or acquired resistance to individual drugs [3]. Generally, there are three types of pair-wise drug combination interactions: additive interaction (the effect (e.g., tumor cell growth inhibition) of a drug combination is equal to the sum of effects of individual drugs taken separately), synergistic interaction (the effect of a drug combination at specific doses is greater than the sum of effects of individual drugs) and antagonistic interaction (the effect of a drug combination is less than the sum of effects of individual drugs). A few synergistic drug combinations have been discovered, for example, targeted therapy and immune treatment [4], targeted therapy and targeted therapy [5], targeted therapy and chemotherapy combinations [6]. Also mechanisms of synergy of drug combinations are diverse and distinct, for example, two drugs can achieve synergy by targeting different oncogenic pathways or mutations simultaneously [7], [8].

It is a challenging task to experimentally screen drug combinations considering the enormous number of possible combinations and heterogeneous mechanisms of synergy. Therefore, computational models to predict effective drug combinations for specific cancer subtypes or individual patients is important and urgently needed. To facilitate the development of computational models of drug combinations, more and more datasets of drug combinations have been being generated. For example, 178 drug combinations were collected from FDA orange book by Zhao et al. [9], and 239 drug combinations were derived from literature curated by Huang et al. [10]. The DCDB database [11] has 1363 drug combinations, and the ASDCD database [12] has 210 drug combinations. Moreover, experimental screening of combinations of 39 drugs against 38 cell lines were reported in [13]; and screening results of combinations of 104 drugs against 59 cell lines were reported in [14].

In spite of the existing approaches, drug combination prediction is still an open problem. For example, Li et al. [15] developed a network propagation method to predict drug synergy based on the gene-gene interaction network. Janizek et al. [16] used XGBoost method combing with a tool named TreeSHAP to predict and make prediction results explainable. A system biology method was designed by Feala et al. [17]; Nelander et al. [18] built a nonlinear multiple-input, multiple-output network model. Chen et al. [19] predicting potential combinations based on semi-supervised learning. Also some computational models [20],[21],[22] have been proposed integrating the genomics profiles of tumor and reverse gene signatures of drugs derived from connective map database [23],[24]. Deep learning models have been developed to solve medical problems and outperformed the traditional machine learning methods. For example, Preuer et al.

[25] proposed a deep neural network (DNN) model named DeepSynergy with an improvement of 7.2% over Gradient Boosting Machines for synergy task. Chiu, Y. C et al. [26] created a model using DNN to predict single drug response of tumors. Gao et al. [27] predicted drug target using a recurrent neural network and convolution neural network. However, the large number of features used in deep learning models might cause overfitting problem. Also, most models were trained using the genomics data of a small number of cancer cell-lines, and it is hard to generalize the model to new cancer cell lines or patients. Moreover, most deep learning models were not explained, and it is hard to understand the potential mechanism of synergy of drug combinations.

To improve the prediction accuracy and explain the deep learning models (black box), in this study, we proposed a novel model, **AuDNNsynergy (Deep Neural Network Synergy model with Autoencoders)**, for synergistic drug combination prediction. The comparison results indicated that the AuDNNsynergy model outperformed four state-of-art approaches in ten model measurements. In this model, multi-omics datasets, i.e., gene expression, mutation, copy number variation, and physicochemical properties of drugs were integrated to achieve better prediction accuracy. Moreover, to make the model has prediction ability for unseen cell lines and tumors, we trained autoencoders with all tumor samples form The Cancer Genome Atlas (TCGA) (pan-cancer dataset). Also, we used a particular cross-validation technique to select the best hyperparameters for the deep learning model. Finally, we conducted the interpretation analysis of the deep learning model to discover potential driver genetic features and uncover underlying mechanisms of synergistic drug combinations.

## 2. Materials
### 2.1 Data
**Drug combination screening dataset.** The high-throughput drug combination screening dataset was downloaded from [13]. In specific, 583 pairwise drug combinations of 38 individual drugs were screened against 39 human cancer cell lines. The 39 cell lines include seven types of cancers, i.e., lung, breast, skin, large intestine, pleura, prostate, and ovary. Tumor cell viability after 48 hours of drug treatments (4 doses for each drug) were measured to estimate the synergy of drug combinations. In total, there are 23062 data points.

**Omics data of cancer cell lines and TCGA samples.** RNA-seq data of the 1156 cancer cell lines and 10,535 tumor samples from The Cancer Genome Atlas (TCGA) were downloaded from Cancer Cell Line Encyclopedia (CCLE) and UCSC Xena Server [28] (URL: https://xenabrowser.net/datapages/?dataset). The somatic mutation data of the screened cell lines and 9104 TCGA tumor samples were downloaded from Catalogue of Somatic Mutations in

Cancer (COSMIC) [29] and UCSC Xena respectively. The copy number of 10,845 TCGA pan-cancer tumors and the 39 cancer cell lines can be accessed from UCSC Xena and CCLE respectively. Some cancer cell lines data cannot be found in CCLE, and COSMIC. We used the genomics data of EFM-129A and COLO320 for the EFM-129B and COLO320DM cell lines respectively. In brief, the EFM-129B established from the same person of EFM-129A after 14 days. The COLO320DM is children of COLO320.

**Physicochemical features of drugs.** To characterize the chemical structures of individual drugs, we used the counts of extended connectivity fingerprints with a radius of 6 (ECP_6) [30], predefined physicochemical properties and presence or absence of toxicophores substructures, which were obtained by using jCompoundMapper [31] and Chemopy [32] respectively.

## 2.2 Data pre-processing

The gene expression data was normalized as $\log_2(tpm + 0.001)$, where $tpm$ denotes the number of transcripts per million. The mutation data (only non-silent mutations were used) was formalized as a binary variable: 1 (with mutation) and 0 (no mutation). Copy number variation (CNV) were estimated using GISTIC2 method [33]. For synergy score of drug combinations, the Loewe Additivity model was used[34]. All features were filtered by removing elements with zero variance and then normalized to has zero-mean and unit variance. We denote $E^T$, $E^C$, $M^T$, $M^C$, $C^T$ as processed data, where $E$, $M$, and $C$ represent gene expression, mutation, copy number and drug, and the T and C indicate tumor and cell lines. For example, $E^T$ presenting TCGA expression data after pre-processing. After preprocessing, each cell line contains 50196, 17643 and 20318 features of expression, mutation, and copy number, and each drug represented using the 2431 physicochemical features, denoted by $D$. In total, each drug combination against a given cancer cell line has 93019 features.

## 3. Model
### 3.1 Architecture of AuDNNsynergy

The AuDNNsynergy model consists of two major components: three autoencoders and a deep neural network, as shown in

**Figure *1*.**
**Settings of autoencoders.** The three autoencoders are the expression autoencoder ($En_{ex}$), mutation autoencoder ($En_{mu}$) and copy number autoencoder ($En_{co}$), which were trained using

TCGA multi-omics data to transfer the knowledge embedded in the large-scale genomics data of TCGA samples. The cell lines' gene expression, copy number and mutation data are then encoded using the trained three autoencoders, and the outputs are used as the input of the prediction network. In specific, we applied six densely connected layers neural network structure for each autoencoder. Moreover, to choose the optimal autoencoder structures and parameters, the number of neurons and batch size were set as hyperparameters, which result in 20 alternatives of autoencoders. The shape of layers is conic and symmetry with five hidden layers have the number of neurons selected from [8192, 4096, 2048, 1024, 512], and the batch size selected from [512, 256]. The linear activations were used in the output layer and bottleneck layer, and the rectified linear activations were used for other layers. The minimum square error (MSE) was used as the cost function minimized by Adam optimizer with default learning rate. The He's uniform distribution was used for initialization, and a moving average over five epochs on a validation set early-stopping was used for training iterations.

**Settings of the deep neural network**. The shape of layers is conic. The number of layers, the number of neurons and learning rate were set as hyperparameters. The number of layers was selected from [2,3], the number of neurons was selected from [8192, 4096, 2048, 1024, 512]; and the number of learning rate was selected from [$10^{-1}, 10^{-3}, 10^{-5}$]. The linear activations were used in the output layer, and rectified linear activations were used in the hidden layers. The batch size of 64, a dropout rate of 0.2 and 0.5 for the input and hidden layers, and stochastic gradient descent optimizer with momentum 0.5 to minimizes MSE cost function were used. The He's uniform distribution was used for initialization, and a moving average over 15 epochs on a validation set early-stopping was used for training iterations. Mathematically, the output $a^i$ of layer $i$ is calculated by $a^i = A^i(W^i a^{i-1} + b^i)$, where $a^i$ denotes the production of layer $i$ ($a^0$ is the input), $A^i$, $W^i$ and $b^i$ are the activation function, weight matrix and bias vector of layer $i$ respectively. The weights and biases are learned by backpropagation to minimizes the MSE cost function.

## 3.2 Cross-validation and model interpretation analysis

For autoencoders' training, each omics data was randomly split into training (90%) and testing (10%) data, and this training was repeated 5 times. For the prediction network, we used five-fold cross-validation experiments to estimate the performance of AuDNNsynergy. To make the model robust each drug-drug combination only appeared in one of these five folds. We used one fold of training dataset as validation dataset to tune the model. To interpret the deep learning model, the 'Innvestigate' package [35] was used to analyze networks' predictions to investigate importance of features. The method of pixel-wise explanations by layer-wise relevance propagation (LRP)

[36] was adopted. In specific, we used Innvestigate first to get the important drug and encoded genomic features contributing to the synergy score; and then we used LPR again to the interesting genomic features for the selected encoded nodes of encoders outputs.

### 3.3 Computation environment of AuDNNsynergy

We deployed the model on Google Cloud Platform with system Ubuntu 16.04, four vcpu, 32G memory, and NVIDIA Tesla V100 GPU. The AuDNNsynergy was run under CUDA 9.0, cuDNN 9.0 using Keras 2.1 with GPU supporting Tensorflow 1.4 backend.

### 3.4 Model comparison

We compared AuDNNsynergy with four state-of-arts machine learning models, including DeepSynergy [25], Gradient Boosting Machines [37], Random Forests [38], and Elastic Nets [39]. We evaluated these model, except DeepSynergy, on the 8846 features dataset [25] and on the 93019 features dataset constructed in this study. The DeepSynergy model was only evaluated on the 8846 features dataset. The reason is that the memory of NVIDIA Tesla V100 GPU is not enough to train it using the 93019 features dataset, which is a limitation of DeepSynergy. To evaluate the DeepSynergy model (a three-layer deep neural network), top 3 hyperparameter settings of performance were used. For the gradient boosting ensembles model, the number of trees and number of features in each split were set as hyperparameters. Random decision forests are also an ensemble machine learning method operated by constructing a multitude of decision trees. Different numbers of trees and numbers of features in each split were evaluated. The Elastic Net is a regularized regression method with a linear combination of lasso and ridge as a penalty of sparse constraint. We performed it with different $\alpha$ values and $L_1$ ratio that controls the relative weights of $L_1$ and $L_2$. The hyperparameter settings of these four methods can be found in **Table 1.**

### 4. Results

### 4.1 Synergistic drug combination prediction

The $En_{ex}$ autoencoder with 8192, 2048 and 512 neurons in three layers respectively, and batch size 256; The $En_{ex}$ autoencoder with 8192, 2048 and 1024 neurons in three layers respectively and batch size 256, and the $En_{ex}$ autoencoder has 8192, 2048 and 1024 neurons in three layers and batch size 256, have the best performance. The two-layers deep neuron network with 8192 and 2048 neurons respectively and $10^{-5}$ learning rate has the best performance. The performance of the models was measured using the rank correlation and MSE measurements between the experimental synergy scores and the predicted synergy scores. The rank correlation coefficient measures the rank correlation for sometimes predicting priorities of drug pairs is more important than accurately predicting the sensitivity to all potential drugs, and MSE is commonly used to assess the quality of an estimator. The prediction results, in terms of rank correlation and MSE, of the proposed **AuDNNsynergy** model are shown in **Figures 2**, **3**, **4** and **5**. As can be seen, the prediction accuracy varies significantly among cell lines and drugs, which might be caused by the distinct mechanisms of different drug-drug pair synergistic/antagonistic effects on specific cell lines. For individual cell lines, the rank correlation coefficients vary from 0.56 to 0.81. Only three cell lines exhibit a correlation below 0.6, and >55% of the cell lines can be predicted with a correlation >0.7. The MSE varies in the range of 61 to 526. Similarly, only three cell lines have MSEs >400; and 50% of the cell lines have MSE < 0.7. For individual drugs, the rank correlation coefficients fall in the range of 0.55 to 0.86. Only two drugs have rank correlation coefficients below 0.6; and ~67% of the drugs have rank correlation coefficients higher than 0.7. The MSEs for drugs vary from 64 to 846. Three drugs have MSEs larger than 400; and ~ 67% of the drugs have MSEs lower than 200. Specifically, the Sunitinib and Doxorubicin are relatively easy to predict, with rank correlation values above 0.7 and MSE below 100. Moreover, we ranked observed synergy scores and compared with the drug pairs with the highest predicted scores. For over 93% test cell lines, the most synergistic drug pair predicted by AuDNNsynergy falls in the top 5 synergistic drug combinations based on the experimental data. For over 74% cell lines, the model correctly predicted the most synergy pairs. We summarized this result using the pie chart, as shown in **Figure 6**.

We also investigated the poor predictions, which is defined as the difference between prediction synergy value and experimental value is over 100. There are 51 samples (2‰ of all data points) in total, and 10 of them are from the LNCAP (prostate adenocarcinoma) cell line; 9 of them are from the NCIH2122 (lung cancer) cell line; and 12 of them are from the NCIH23 cell line. For drugs, the combinations of MK-4827, GELDANAMYCIN and MK-8776, BEZ-235 appear most frequently with six times in the poor predictions. In summary, most of the predictions are good, as shown in the scattering plot in **Figure 7**.

**4.2 Model comparison**
The detailed model comparison results are listed in **Table 2** and **Table 3**, using the measurements of rank correlation, MSE, confidence interval, root mean square error (RMSE) and Pearson correlation. The models were evaluated on the 93019 features dataset, and the 8846 features dataset (*marked). As can be seen, the proposed AuDNNsynergy model outperforms other four models, especially in terms of the rank correlation (0.73) and MSE (241.12). Moreover, it can be seen that other models could not improve their performance even using more features. Also, it is hard to apply the DeepSynergy directly on the 93019 features dataset because there will be more than six hundred million parameters to be optimized. Therefore, it cannot be run on the google cloud platform. **Table 3** shows the area under the receiver operator characteristics curve (ROC AUC), area under the precision-recall curve (PR AUC), accuracy (ACC), precision (PREC) and Cohen's Kappa of the five models. The proposed AuDNNsynergy model has the best performance in terms of ROC AUC, PR AUC, ACC, PREC and Kappa compared with other models.

## 5. Interpretation analysis
### 5.1 Important omic features for the overall prediction

Using the interpretation analysis, the number of most important features that can negatively regulate, positively regulate and impact the model are listed in **Table 4**, **Table 5**, and **Table 6**, respectively. As can be seen, over 50% of most important features are the chemical structures. It is interestingly that the number of important omics features almost will not change with more feature selected. To identify the important genes, we employ the interpretation analysis twice. First, we used the interpretation analysis to get the encoded omics features in the top 100 important features; and repeat five times to select the common features. Then we conduct the interpretation analysis again on the autoencoders to identify the related genes. In summary, the gene expression of 2 genes, mutation of 25 genes, and copy number of 14 genes were selected, which are listed in **Table 7**. Through the literature review, we found that many of these genes have been reported related to drug resistance or sensitivity, e.g., STAT3, miR-222, XIAP, FGF16, PBRM1, and SPP1. For example, the feedback activation of STAT3 is a common cause of resistance to many targeted cancer therapies and chemotherapies [40]. The miR-222 confers Adriamycin and Docetaxel resistance in MCF-7 cells [41]. The FGF16 gene regulates aggressiveness of HepG2 and PLC/PRF/5 cells [42]. The mutation of PBRM1, a subunit of the SWI/SNF complex, attenuated the effect of Gefitinib in part by sustaining AKT pathway function during EGFR inhibition [43]. The SPP1 gene plays an important role in drug sensitivity of malignant pleural mesothelioma [44], Oral Squamous Cell Carcinoma [45], and pancreatic cancer [46]. SPP1 is also pivotal in tumor cell proliferation, angiogenesis and metastasis in ovarian [47], brain [48], lung [49], kidney [50], liver [51], bladder [52], breast [53], esophageal [54], gastric [55], colon [56] and prostate cancers [57].

### 5.2 Mechanism of synergy for specific drug combinations against given cell lines

We also explored underlying mechanisms of synergy of some drug combinations. For example, Dasatinib and Carboplatin have synergy effect on ovarian cell line A2780 [58]. Dasatinib targets the SRC while carboplatin targets the DNA. We found the most important gene is the copy number of PTPRK, which dephosphorylates SRC [59]; Drug combination Sorafenib and Erlotinib are reported to treat non-small cell lung cancer (NSCLC) [60]. Erlotinib and Sorafenib are the inhibitors of epidermal growth factor receptor (EGFR) and vascular endothelial growth factor (VEGF) receptor respectively. The two of the most important omics predictors of this response are gene PAIP2 and gene miR-507. PAIP2 has a positive effect on the response while miR-507 has an adverse effect. Suppressing miR-507 promotes VEGF-C production [61], and PAIP2 is a strong regulator of VEGF [62]. However, the PAIP2, miR-507, and PTPRK gene are not in the most critical predictors for all drug combination predictions. The results indicate that the interpretation analysis could be useful for uncovering novel mechanism of synergy for specific drug combination against given cancer cell-lines.

**5.3 Synergistic drug combination prediction for TCGA samples**

We also applied the AuDNNsynergy to predict synergy score of the 853 drug combinations on TCGA samples. Although the omics profiles of cell lines and tumors might be different, some prediction results are still meaningful. For example, drug combination Dasatinib/Erlotinib has a high synergy score (rank 7/583) for the NSCLC bears EGFR mutation. Dasatinib is an inhibitor of SRC while Erlotinib is EGFR inhibitor and this pair has been reported are synergy, safe and feasible in NSCLC with activating EGFR mutations [63]. Also, the Dasatinib with another EGFR inhibitor, Laptinib, have a high synergy score (rank 5/583) for NSCLC with EGFR mutation. The other example is the ABT-888/BEZ-235, MK-4827/BEZ-235, and MK4827/MK8669 combinations, which are top ranked in our prediction (top 4%-7%). ABT-888 and MK-4827 target on PARP while BEZ-235 and MK-8669 target on MTOR on breast cancer samples. The literature review [64] showed that the treatment of a BRCA2 mutated luminal breast cancer PDX by Everolimus and Olaparib combination resulted in tumor regression. Interestingly, Olaparib is a PARP inhibitor while Everolimus is a mTOR inhibitor, and their combination could lead to unrepaired DNA damage and tumor regression *in vivo*, through a cross-talk between DNA repair and mTOR pathways. These promising examples show that the proposed model can be used to predict drug combinations for new tumor samples not in the 39 cell lines in the screening data. One of the reasons is that the proposed AuDNNsynergy model is trained using the TCGA samples, and it can transfer and maintain the knowledge of different tumor samples to achieve robust predictions for new tumor samples.

## 6. Discussion

In this study, we have presented a deep learning model, AuDNNsynergy, which outperforms existing models of drug combination prediction. In the model, we integrated multi-type of genomics data, e.g., gene expression, copy number and mutation using three autoencoders; and applied the transfer learning to train the autoencoders using TCGA data. The results indicated that the transfer learning of TCGA data can improve the prediction accuracy. Also, we conducted the interpretation analysis, and identified a set of interesting omics features for drug combination prediction, and uncovered potential mechanisms of synergy of specific drug combination against given cancer cell lines.

In the future work, we aim to improve the performance of AuDNNsynergy as follows. 1) A larger drug combination screening dataset will be used to evaluate the AuDNNsynergy model. There are a few drug combination datasets reported. However, the measurements of drug combination synergy are heterogeneous. We will merge these screening datasets using a standard measurement of combinations effectiveness to evaluate the AuDNNsynergy model. 2) More types of genomics data, e.g., methylation, PARADIGM pathway activity, protein expression, and drug structure graph can be added to improve the prediction model.

**Table 1.** Hyperparameter settings of 4 method.

| Method | Hyperparameter |
|---|---|
| Deep Synergy | $number\ of\ hidden\ layers\ neurons \in \{[8192,4096],[8192,8192],[8192,2048]\}$ |
| Gradient Boosting Machines | $number\ of\ estimators \in [128,256,512,1024]$<br>$features\ considered \in [log_2(\#\ of\ features), square(\#\ of\ features)]$ |
| Random Forests | $number\ of\ estimators \in [128,256,512,1024]$<br>$features\ considered \in [log_2(\#\ of\ features), square(\#\ of\ features)]$ |
| Elastic Nets | $\alpha \in [0.1, 1, 10, 100]$<br>$L_1\ ratio \in [0.25, 0.5, 0.75]$ |

*Table 2*. Rank Correlation, MSE, Confidence Interval, RMSE and Pearson correlation for 5 methods on the 93019 features dataset, and the 8846 features dataset (*marked).

| Method | Rank Correlation | MSE | RMSE | Pearson correlation |
|---|---|---|---|---|
| AuDNNsynergy | 0.73±0.02 | 241.12±43.52 | 15.46±1.44 | 0.74±0.03 |
| Deep Synergy* | 0.66±0.02 | 255.49±49.54 | 15.91±1.56 | 0.73±0.04 |
| Gradient Boosting Machines | 0.64±0.02 | 310.73±50.40 | 17.57±1.42 | 0.64±0.02 |
| Gradient Boosting Machines* | 0.64±0.01 | 308.06±49.34 | 17.50±1.39 | 0.64±0.02 |
| Random Forests | 0.64±0.02 | 326.31±53.54 | 18.00±1.45 | 0.63±0.02 |
| Random Forests* | 0.63±0.02 | 330.18±51.45 | 18.12±1.37 | 0.61±0.03 |
| Elastic Nets | 0.49±0.02 | 422.72±54.35 | 20.52±1.30 | 0.44±0.02 |
| Elastic Nets* | 0.49±0.02 | 424.38±54.39 | 20.56±1.30 | 0.44±0.02 |

**Table 3**. ROC AUC, PR AUC, ACC, PREC and Kappa for 5 methods.

| Performance Metric | ROC AUC | PR AUC | ACC | PREC | Kappa |
|---|---|---|---|---|---|
| AuDNNsynergy | 0.91±0.02 | 0.63±0.06 | 0.93±0.01 | 0.72±0.06 | 0.51±0.04 |
| Gradient Boosting Machines | 0.87±0.02 | 0.54±0.04 | 0.93±0.01 | 0.72±0.03 | 0.39±0.05 |
| Random Forests | 0.86±0.03 | 0.51±0.04 | 0.92±0.02 | 0.57±0.04 | 0.21±0.04 |
| Elastic Nets | 0.77±0.04 | 0.33±0.09 | 0.92±0.01 | 0.23±0.29 | 0.15±0.09 |
| Deep Synergy* | 0.90±0.03 | 0.59±0.06 | 0.92±0.03 | 0.56±0.11 | 0.51±0.04 |

**Table 4**. The number of 4 types features of most positive effect for synergy scores.

| Features | 100 | 500 | 1000 | 2000 |
|---|---|---|---|---|
| Expression | 0.2±0.4 | 0.2±0.4 | 0.2±0.4 | 0.2±0.4 |
| Mutation | 9.4±0.5 | 9.6±0.8 | 9.8±1.2 | 0.6±0.8 |
| Copy Number | 11.2±2.0 | 12.2±2.3 | 12.4±2.2 | 12.6±2.1 |
| Drug | 79.2±1.6 | 478.0±1.8 | 977.6±1.3 | 1976.6±1.2 |

**Table 5**. The number of 4 types features of most negative effect for synergy scores.

| Features | 100 | 500 | 1000 | 2000 |
|---|---|---|---|---|
| Expression | 1.8±0.4 | 1.8±0.4 | 1.8±0.4 | 4.6±5.2 |
| Mutation | 11.8±0.4 | 12.2±0.7 | 12.2±0.7 | 36.2±48.4 |
| Copy Number | 13.2±1.7 | 13.4±1.9 | 13.6±2.2 | 143.8±97.0 |
| Drug | 73.2±1.2 | 472.6±1.5 | 972.4±1.6 | 1820.8±75.1 |

**Table 6.** *The number of 4 types features effect synergy scores most.*

| Features | 100 | 500 | 1000 | 2000 |
|---|---|---|---|---|
| Expression | 2.0±0.0 | 2.0±0.0 | 2.0±0.0 | 2.0±0.0 |
| Mutation | 20.8±0.4 | 21.4±0.5 | 21.8±0.7 | 22.2±0.4 |
| Copy Number | 24.2±0.7 | 25.2±0.7 | 25.4±0.8 | 25.8±0.4 |
| Drug | 52.8±0.4 | 451.4±1.0 | 950.8±1.2 | 1950.0±0.0 |

**Table 7.** Important genomic features.

| Feature type | Gene symbol |
|---|---|
| Expression | miR-222, STAT3 |
| Mutation | HYDIN, MUC19, DNAH3, TBCE, MDN1, KIF21B, SHARPIN, RIBC2, NEB, USP53, USP35, SPP1, PCDH1, KLF5, ANK3, CMTM1, PPL, STK32B, C21orf67, LRP1B, DMD, AIM2, NAA60, XRCC6BP1, PPFIA2 |
| Copy number | PRPF39, SNORD22, PBRM1, SNORA68, SCARNA14, SNORD23, MIRLET7G, SNORA77, INO80D, XIAP, CNOT10, FGF16, LCTL, SNORA43 |

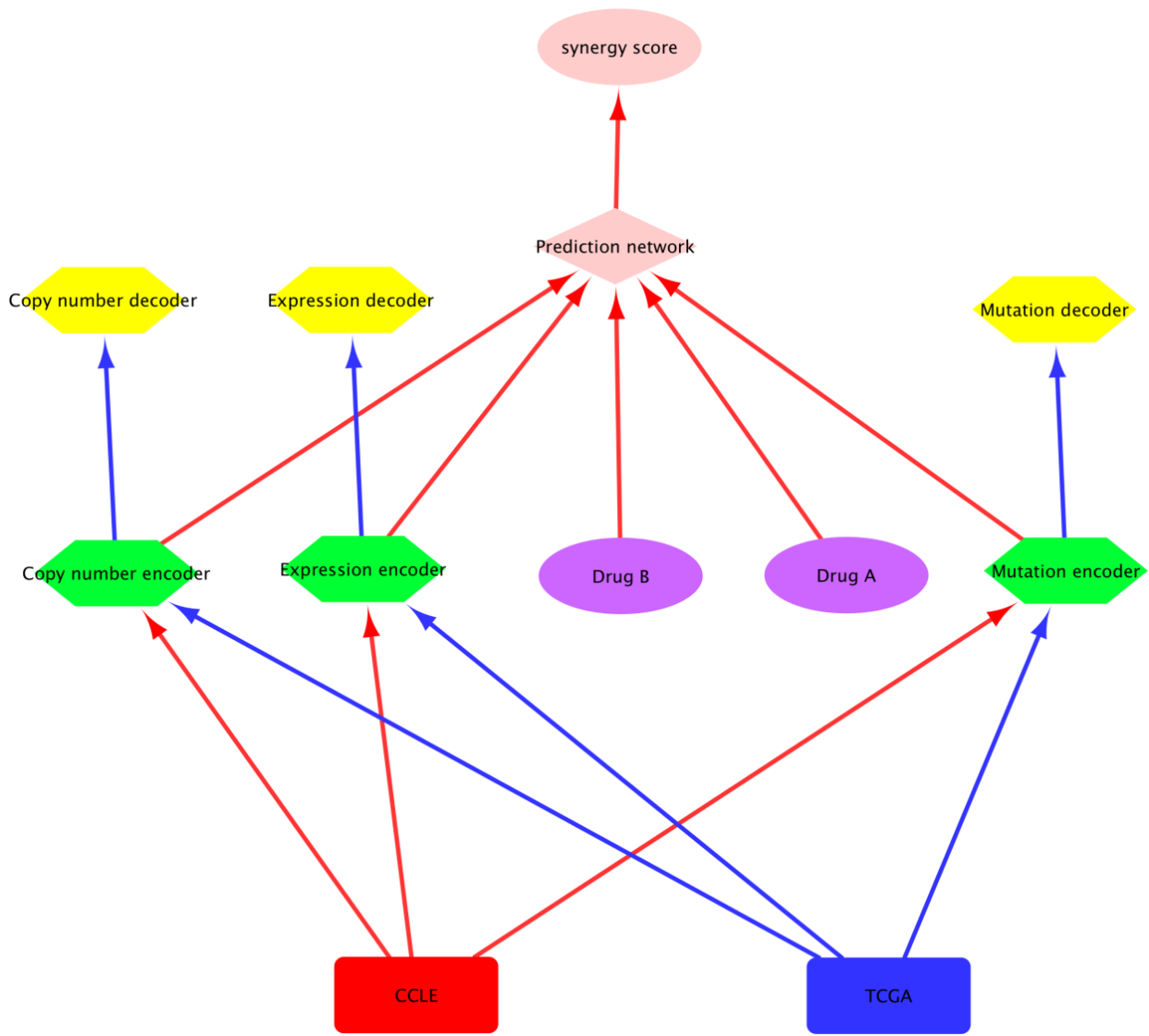

**Figure 1**. Overview of the AuDNNsynergy architecture.

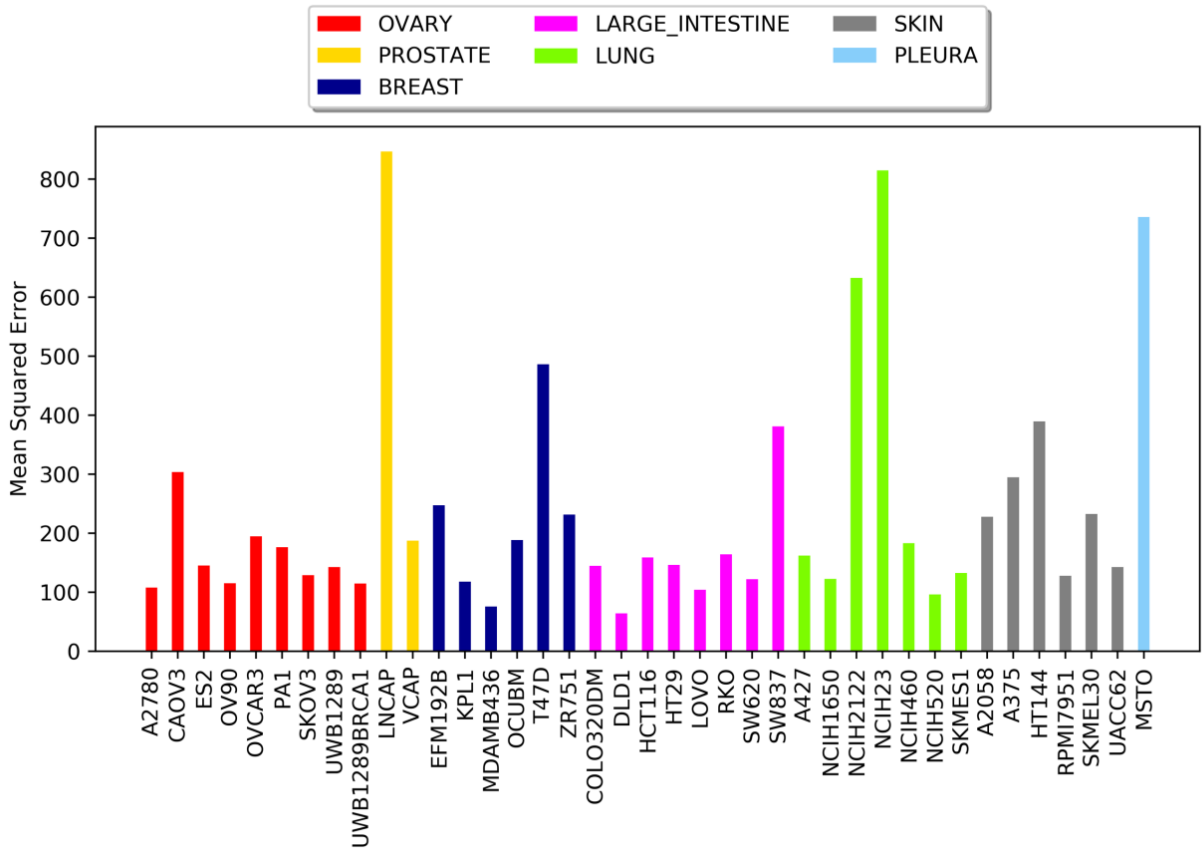

**Figure 2**. MSE of predicted synergy scores per cell line.

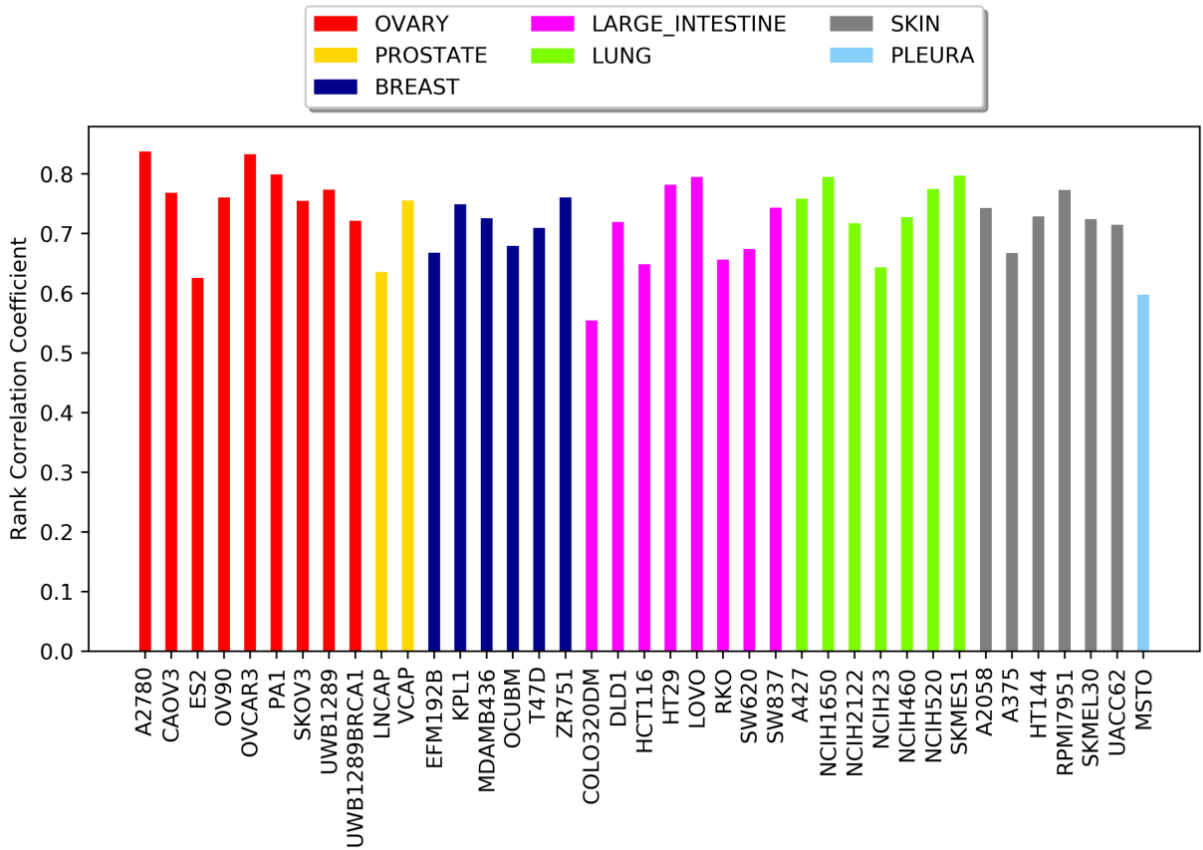

**Figure 3.** Rank correlation coefficients between observed and predicted synergy scores per cell line.

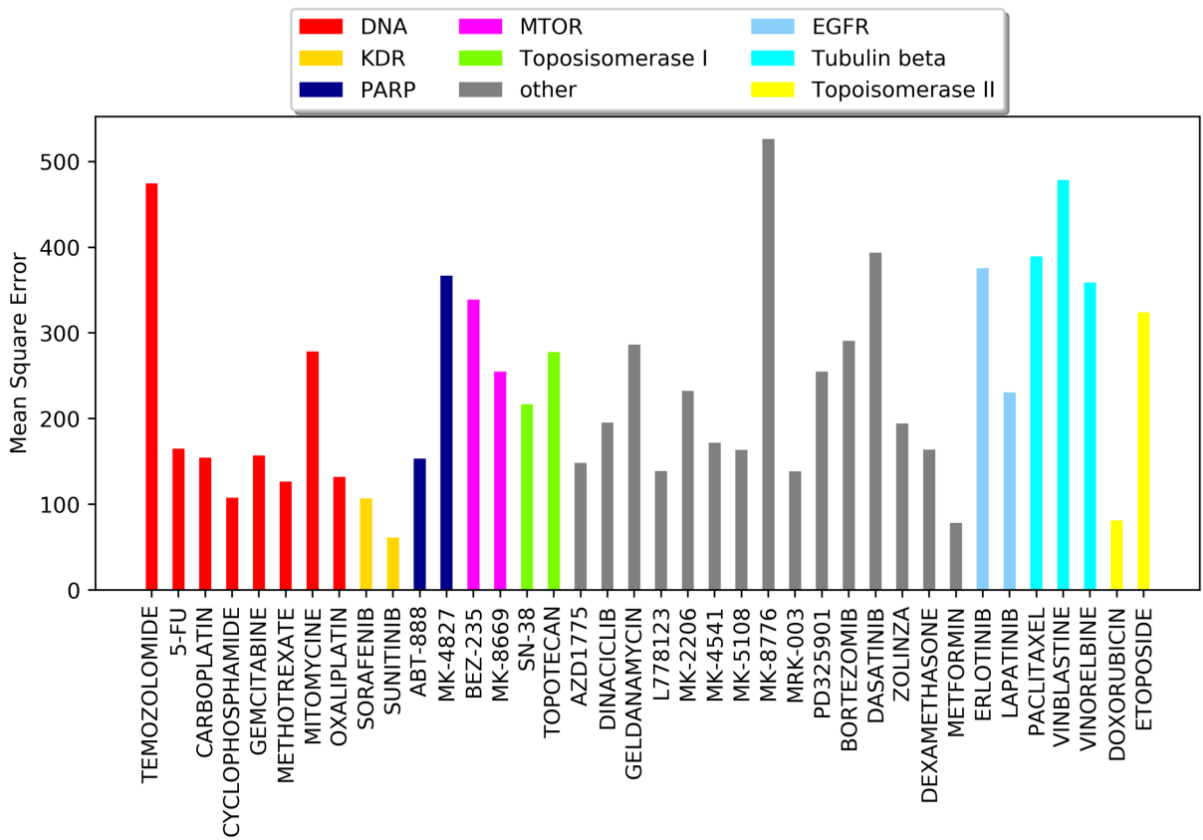

**Figure 4.** MSE of predicted synergy scores per drug.

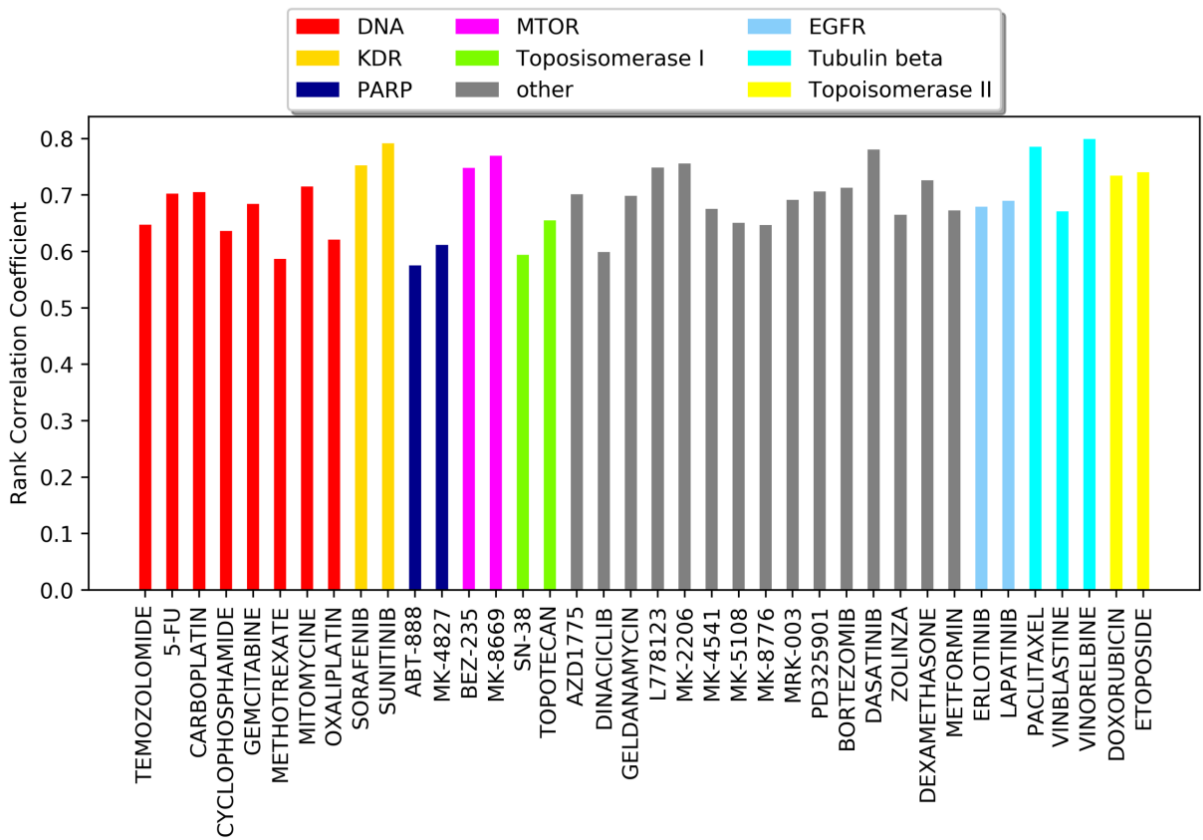

**Figure 5**. Rank correlation coefficients between observed and predicted synergy scores per drug.

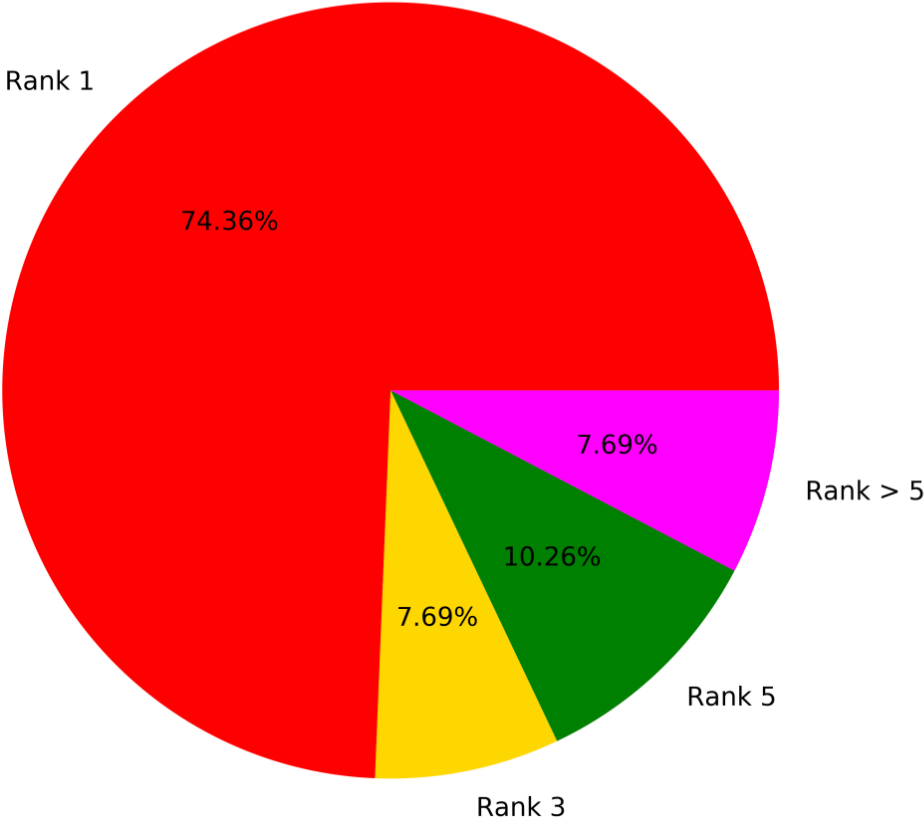

**Figure 6.** Ranking correlation between prediction and experimental validations.

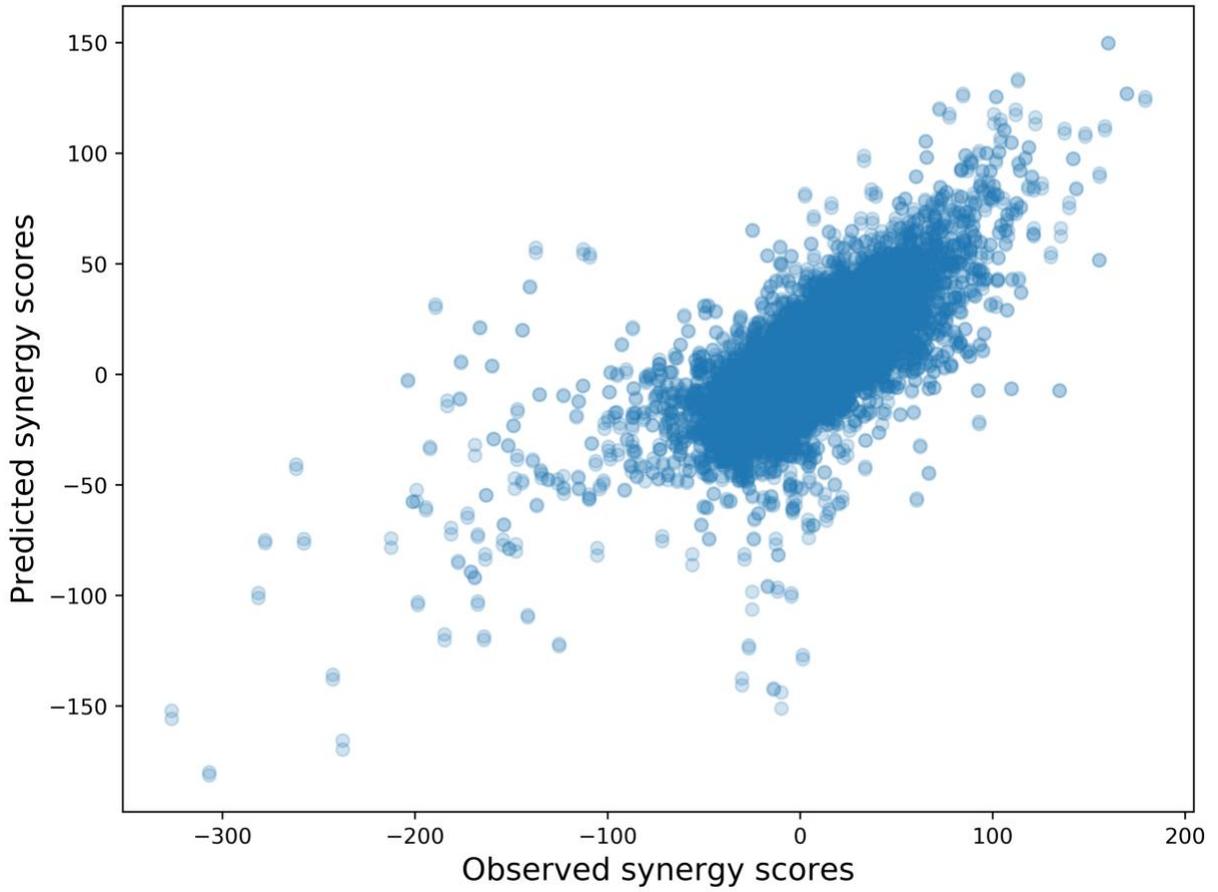

**Figure 7.** The overall distribution of predicted and the experimental validated synergy scores.